\documentstyle[multicol,aps,prl,epsf]{revtex}
\begin{document}
%\draft
\title{Transport coefficients of a mesoscopic fluid dynamics model}
\author{N. Kikuchi, C. M. Pooley, J. F. Ryder and J. M. Yeomans}
\address{Department of Physics, Theoretical Physics, 1 Keble Road,
Oxford, OX1 3NP, England.
}
\date{\today}
\maketitle
%--------------------------------------------------------------------------
\begin{abstract}
We investigate the properties of stochastic rotation dynamics, a mesoscopic model used for simulating fluctuating hydrodynamics.  Analytical results are given for the transport coefficients. We discuss the most efficient way of measuring the transport properties and obtain excellent agreement between the theoretical and numerical calculations.
\end{abstract}

\begin{multicols}{2}
%--------------------------------------------------------------------------
\section{Introduction}

Investigating the hydrodynamics of complex fluids is a challenging task because of the large number of degrees of freedom involved, the interplay between the flow and thermodynamic properties and the existence of widely varying length and time scales. Examples include the dynamics of polymers and biopolymers in solution, polymer melts, amphiphilic systems, particulate suspensions and chemically reacting fluids.

The complexity of these systems points to numerical approaches and it has become apparent that mesoscale algorithms such as dissipative particle dynamics\cite{GW97}, smoothed particle hydrodynamics\cite{JJ92} and lattice Boltzmann algorithms\cite{S01,CD98} are increasingly providing useful numerical tools to investigate complex fluids and, in particular, the role of hydrodynamics in their behaviour. These approaches are essentially novel ways of solving the Navier-Stokes equations designed to incorporate the most relevant microscopic physics of a given complex fluid.

Recently Malevanets and Kapral described one such mesoscopic approach which we shall term stochastic rotation dynamics\cite{MK99}. This method solves the thermohydrodynamic equations of motion by following the paths of particles moving in discrete time but continuous space. The algorithm has two major advantages. Firstly, because it is a particulate method, it is easy to couple it to a solute, say, polymers or colloids, which can be treated using a molecular dynamics algorithm\cite{MK00,MY00,KGY02,FPVA03}. Secondly, the algorithm is rather simple and therefore analytical work is possible which can be aimed at increasing our understanding of mesoscopic approaches to the hydrodynamics of complex fluids.

Malevanets and Kapral mapped the stochastic rotation algorithm to a Boltzmann equation and showed that the model possesses an H-theorem\cite{MK99}. Ihle and Kroll extended the algorithm to correct for spatial correlation effects and gave approximate results for the viscosity and diffusion coefficient in two dimensions\cite{IK01}. Allahyarov and Gompper\cite{AG02} used a Poiseuille flow geometry to mesure the viscosity in three dimensions. Stochastic rotation dynamics has been applied to colloids in solution\cite{MK00}, dilute polymer solutions\cite{MY00,KGY02,FPVA03}, amphiphilic systems\cite{HCO00}, flow in porous media\cite{S02}, binary fluid mixtures\cite{MY00b}, cluster dynamics\cite{LK01} and flow around solid objects\cite{LGIK01a,LGIK01b,ICO02}.

To understand as fully as possible how stochastic rotation dynamics can be used to simulate hydrodynamic behaviour it is helpful to increase our understanding of its properties. Therefore this paper presents an investigation of the viscosity and diffusion coefficient of the model in two and three-dimensions. 

The algorithm is described in section $2$ following Malevanets and Kapral\cite{MK99} and Ihle and Kroll\cite{IK01}. We then show how it is possible to model a shear flow by imposing Lees-Edwards boundary conditions\cite{LE72,AT89}. The ability to simulate shear is important as it provides a computationally feasible way of measuring viscosities over a wide parameter range. It will also extend the range of applications of the approach to study viscoelastic effects such as the shear-thinning of a polymer solution\cite{RY03}.

In section $3$ we discuss the shear viscosity, \(\eta\), of stochastic rotation dynamics. We argue that it is useful to divide the viscosity into two contributions, a kinetic and a collisional term. We first provide analytical expressions for the kinetic and collisional parts of the viscosity. We then compare these to numerical results and discuss their relative magnitudes.

Section $4$ deals with the coupling of a test particle to the solvent model. A massive particle obeys Langevin dynamics and we obtain an expression for the friction coefficient describing the coupling to the solvent. We stress that the friction coefficient is related to the collisional part of the viscosity, not to the total viscosity, a distinction unimportant in a real fluid, but pertinent here.

Finally a conclusion summarises the paper and suggests possibilities for future work.

%--------------------------------------------------------------------------
\section{THE MODEL}
\subsection{Stochastic rotation dynamics}

\par The solvent is modelled by a large number $N$, typically $\sim10^5$, of 
point-like particles of mass $m$ which move in continuous space with a continuous
distribution of velocities, but discretely in time \cite{MK99}. The 
algorithm proceeds in two steps. In the first of these, a free streaming 
step, the positions of the solvent particles at time \(t\), 
\({\mbox{\boldmath $r$}}_i(t)\), are updated simultaneously according to
\begin{eqnarray}
{\mbox{\boldmath $r$}}_i\,(t\!+\!{\delta t})\,=\,{\mbox{\boldmath $r$}}_i\,
(t)+{\mbox{\boldmath $v$}}_i\,(t){\delta t}
\end{eqnarray}
where \({\mbox{\boldmath $v$}}_i(t)\) is the velocity of a particle 
and \({\delta t}\) is the value of the discretized time step of the 
solvent which we take to be unity.

\par The second part of the algorithm is the collision step. The system is 
coarse-grained into \(L^d/a^d\) cells of a $d$-dimensional lattice with side \(a\).  There is no 
restriction on the total number of solvent particles 
in each cell, (although the total number of particles in the system is 
conserved). Stochastic multi-particle 
collisions are performed within each individual cell, by rotating the 
velocity of each particle relative to the centre of mass 
velocity \({\mbox{\boldmath $v$}}_{cm}(t)\) of all the particles 
within that cell
\begin{eqnarray}
{\mbox{\boldmath $v$}}_i\,(t\!+\!{\delta t})\,=\,{\mbox{\boldmath $v$}}_{cm}
(t)+{\mbox{\boldmath $R$}}\left(\,{\mbox{\boldmath $v$}}_i\,(t)-
{\mbox{\boldmath $v$}}_{cm}(t)\,\right)\label{COLLISIONSTEP}.
\end{eqnarray}
\({\mbox{\boldmath $R$}}\) is a rotation matrix which rotates 
velocities by a fixed angle \(\alpha\) around an axis generated randomly for 
each cell and at each time step.

\par The aim of the collision step is to transfer momentum between the 
particles while conserving the total 
momentum and energy of each cell. Because mass, momentum and energy 
are conserved locally the thermohydrodynamic equations of motion 
are captured in the continuum limit\cite{MK99}. Hence hydrodynamic 
interactions can be propagated by the solvent. Note, 
however, that any molecular details of the solvent are excluded: this 
allows the hydrodynamic interactions to be modelled with minimal 
computational expense.

\par The volume in phase space is invariant under both the free 
streaming and collision steps. Hence the system is described by 
a microcanonical distribution at equilibrium. Also it has 
been proved that the algorithm obeys an H-theorem\cite{MK99}.
 
\par A particularly useful feature of stochastic rotation dynamics 
is the ease with which hydrodynamic interactions can 
be ``turned-off'' thus replacing the hydrodynamic heat bath by a 
Brownian (random) heat bath. This is achieved by randomly 
interchanging the velocities of all the solvent particles after 
each collision step, thus relaxing the constraint of  
momentum conservation from a local to a global one. Accordingly the velocity 
correlations which result in hydrodynamic interactions disappear 
from the fluid although the equilibrium Maxwell-Boltzmann distribution
is maintained. This greatly facilitates pinpointing the effects of the 
hydrodynamic interactions in a given situation\cite{KGY02}.

\subsection{Simulation parameters}

\par Unless otherwise stated $m=1$, $a=1$ and the number of cells was $L^3=32^3$.
The initial solvent distribution was generated by assigning positions 
randomly within the system with an average number of particles per unit cell \(\gamma=5\). Thus the total number of particles was
$N=163,\!840$. The velocities were assigned from a uniform distribution 
\(\left(-v_{max}{\leq}v_{\alpha}{\leq}v_{max}\right)\),  
\(\alpha=x,y,z\) where \(v_{max}=\sqrt{\frac{3k_BT}{m}}\). The
distribution relaxed rapidly (\(t{\sim}100\)) to 
the equilibrium Maxwell-Boltzmann form corresponding to the temperature
$T$. Because of the long time-scales studied, any small non-zero net momentum in the system can eventually cause a large shift of a velocity distribution along the \(y\) axis. To overcome this we perform a Galilean transformation to remove any net momentum from the system and then rescale the velocities to restore the correct temperature. 

\subsection{Lees-Edwards shear boundary condition}

\par To achieve a steady shear flow we use Lees-Edwards boundary 
conditions\cite{LE72,AT89}, which are rather naturally incorporated 
into stochastic rotation dynamics. Our aim is to introduce a flow in the $x$-direction, with a constant velocity gradient along $y$. Periodic boundary conditions are applied along \(x\) and \(z\). Along  \(y\) the boundary conditions are modified so that if a particle  crosses the lower boundary at (\(x, y=0, z\)) it
returns to the upper boundary at the position 
(\(x={\left(x+ut\right)}_{mod.L}, y=L_, z\)) with  
velocity (\(v_x=v_x+u,v_y,v_z\)) where \(mod.L\) denotes 
modulus L. Similarly if a particle leaves the upper 
boundary at (\(x, y=L, z\)), it reappears at the lower boundary at
(\(x={\left(x-ut\right)}_{mod.L}, y=0, z\)) with velocity 
(\({v_x=v_x-u,v_y,v_z}\)). This algorithm gives a steady shear flow, 
and the system mimics one of infinite size.

\par When an external force is applied to the system, energy is constantly 
produced. Moreover a small net momentum can be introduced by the Lees-Edwards boundary condition for any system with a finite number of particles. To sustain a zero net momentum and temperature the appropriate shear velocity is subtracted from each particle velocity, a Galilean transformation is applied, the solvent velocities are rescaled to give a temperature \(T\) and the shear velocity is restored. This procedure is carried out after each collision step.

\subsection{Grid shift for periodic and shear boundary conditions}

\par Ihle and Kroll \cite{IK01} pointed out that at low 
temperatures the transport coefficients of stochastic rotation 
dynamics showed anomalies. This happens because at low temperatures 
particles in a given cell can remain in that cell  over several 
time steps and participate in a series of collisions. 
Accordingly the molecular chaos assumption breaks down.
However Ihle and Kroll \cite{IK01} showed that it is possible to improve the behaviour of the algorithm by placing the cubic grid in a
random position at each collision step.

\par In practice the easiest way to implement such a grid shift is to
move all the particles with the same random vector with components in the interval \(\left[-a/2, a/2\right]\) before each collision step. After the collision step the particles are
returned to their original positions. If the displacement takes the
particles across the edge of the simulation box Lees-Edwards or periodic boundary
conditions are applied as appropriate.

\par We note that the grid shift accelerates momentum transport between cells. As 
a result the system relaxes more quickly. It must also be taken into
account when calculating transport coefficients such as the viscosity.

%--------------------------------------------------------------------------

\section{Viscosity}

\par One of the most important parameters characterising a fluid is its
viscosity. In this section we investigate the shear viscosity associated
with the stochastic rotation model. The viscosity is usefully divided into two contributions. The kinetic
viscosity results from the momentum transferred during the free
streaming step, and the collisional viscosity results from the
momentum transferred during the velocity rotations.

%--------------------------------------------------------------------------

\subsection{Calculation of the kinetic viscosity}

\par To calculate the kinetic viscosity we consider a system undergoing shear with rate \({\dot \gamma}=\frac{{\partial}u_x\left(y\right)}{{\partial}y}\) in the \(x\) direction and use a kinetic theory approach. Consider first two dimensions. On average the velocity profile is given by ${\bf v} = ({\dot \gamma}y, 0)$. In the steady state there is a frictional
force acting on any plane in the fluid perpendicular to $y$  which is
given by the element of the stress tensor 
\begin{eqnarray}
\sigma_{xy}={\eta}\frac{\partial {u_x\!\left(y\right)}}
{\partial y}\label{NONEQSV}
\end{eqnarray}
where \(\eta\) is the shear viscosity. This is the experimental definition of the viscosity\cite{AT89}. Our aim is to calculate the stress tensor during the streaming step: \(\sigma_{xy}=-\)(flux of $x$-momentum crossing a plane of constant $y$).

For simplicity we take \(y=y_{0}=0\). Consider a particle at position (\(x,y\)) with velocity (\(v_x,v_y\)). Particles will only cross the plane during a time step if they are moving towards the plane and have a velocity such that $| v_y{\delta t}|$ is greater than the distance to the plane.
Therefore, the stress tensor can be written
\begin{eqnarray}
\sigma_{xy}=&-\frac{\rho}{\delta t}& \int_{-\infty}^{\infty}  \!\!\!\! dv_x  \int_{-\infty}^{0}  \!\!\!\! dy \int_{-y/\delta t}^{\infty}  \!\!\!\! dv_y \,\, v_x P(v_x - {\dot \gamma} y, v_y) \nonumber \\
+&\frac{\rho}{\delta t}& \int_{-\infty}^{\infty} \!\!\!\! dv_x  \int_{0}^{\infty} \!\!\!\! dy \int_{-\infty}^{-y/\delta t}  \!\!\!\! dv_y \,\, v_x P(v_x - {\dot \gamma} y, v_y)  \nonumber 
\end{eqnarray}    
where \(P(v_x,v_y)\) is the velocity probability distribution of particles in the rest frame of the fluid and \(\rho=\frac{m\gamma}{a^d}\) is the mass density. \(\gamma\) denotes the average number of particles per cell and $d=2,3$ is the dimensionality. By making the change of variable $v_x^\prime = v_x - {\dot \gamma} y$, and changing the order of integration this reduces to
\begin{eqnarray}
\sigma_{xy}=\frac{{\dot \gamma} \rho\,{\delta t}}{2} \left< v_y^2 \right> - {\rho}\left< v_x v_y \right>
\label{sigma}
\end{eqnarray} 
where the averaging is over the probability distribution $P$. The first and second terms result from the shear profile and from induced correlations between $v_x$ and $v_y$ respectively. In the steady state the velocity distribution \(P\) deviates from the Maxwell Boltzmann form because of these correlations.

\par To find the behaviour of  $\left< v_x v_y \right>$ we calculate separately the effect of the streaming and collision operations. Consider the velocity distribution at the plane \(y=y_{0}\). At time \(t\), at \(y_0\), particles with positive velocity \(v_y\) come from \(y_0-v_y{\delta t}\) due to the streaming step, and correspondingly have a smaller $x$ velocity. Conversly particles from \(y>y_0\) tend to have a higher $x$ velocity. This means that the velocity  probability distribution is sheared by the streaming. That is $P^{after}(v_x, v_y) = P^{before}(v_x + {\dot \gamma} v_y \delta t,v_y)$. Averaging $v_x v_y$ over this new distribution and changing variables gives
\begin{eqnarray}
\left<v_xv_y\right>^{after}&=&\int_{-\infty}^{\infty}\!\!\!\! dv_x\int_{-\infty}^{\infty}dv_y\,\,v_xv_y P(v_x+{\dot \gamma}v_y{\delta t}, v_y)\nonumber\\
&=&\left<v_xv_y\right>-{\dot \gamma} \delta t \left< v_y^2 \right>.
\end{eqnarray} 
Thus the streaming operation changes the value of $\left< v_x v_y \right>$ by $-{\dot \gamma} \delta t \left< v_y^2 \right>$, and tends to make $v_x$ and $v_y$ increasingly anticorrelated.

\par Next we consider the effect of the collision step on the velocity distribution. Collisions tend to reduce correlations. To see this we consider a cell of \(n\) particles and the two dimensional collision equation
\begin{eqnarray}
{\mbox{\boldmath $v$}}(t+{\delta t})=\left( \begin{array}{cc} cos \alpha & \pm sin \alpha\\  \mp sin \alpha & cos \alpha \end{array} \right)\left({\mbox{\boldmath $v$}}(t)-{\mbox{\boldmath $v_{cm}$}}\right)+{\mbox{\boldmath $v_{cm}$}}\nonumber
\end{eqnarray}
where the sign corresponds to the direction of the rotation axis. The centre of mass velocity ${\mbox{\boldmath $v_{cm}$}}$ can be divided into two contributions, ${\mbox{\boldmath $v_{cm}$}}=\left({\mbox{\boldmath $v$}}+{\mbox{\boldmath $\hat v$}}\right)/n$, the first from the test particle of velocity ${\mbox{\boldmath $v$}}$ and the second \({\mbox{\boldmath $\hat v$}}=\sum_{i=1}^{n-1}{\mbox{\boldmath $v_i$}}\) from the $n-1$ particles sharing the same cell. We assume molecular chaos, that is the velocities of different particles are uncorrelated. Then, $\left<v_x{\hat v_y}\right>=\left<{\hat v_x}v_y\right>= 0$ and $\left<{\hat v_x}{\hat v_y}\right>=(n-1)\left<v_xv_y\right>$ and
\begin{eqnarray}
& &\left<v_x(t+{\delta t})v_y(t+{\delta t})\right>\nonumber\\
&=&\left[1-\left(\frac{n - 1}{n}\right)\left(1-\cos2\alpha\right)\right]\left<v_x(t)v_y(t)\right>.\nonumber
\end{eqnarray}
where we have averaged over the direction of the rotation axis. In the simulations $n$ is not constant. Therefore, we must consider density fluctuations. The probability of \(n\) particles being in a given cell is given by the Poisson distribution \(P\left(n\right)=\frac{e^{-\gamma}{\gamma}^n}{n!}\). The probability of a given particle being in a cell with a total of $n$ particles is $\frac{nP(n)}{\gamma}$, so taking an average over this distribution
\begin{eqnarray}
& &\sum_{n=1}^{\infty}\frac{nP(n)}{\gamma}\left<v_x(t+{\delta t})v_y(t+{\delta t})\right>\nonumber\\
&=&\left[1-\left(\frac{{\gamma}-1+e^{-{\gamma}}}{\gamma}\right)\left(1-\cos2\alpha\right)\right]\left<v_x(t)v_y(t)\right>\nonumber\\
&=& f\!\left(\alpha,\gamma\right)\left<v_x(t)v_y(t)\right>\label{f}.
\end{eqnarray}

During one time step we find that $\left< v_x v_y \right>$ is first reduced by the streaming operation and then multiplied by a constant factor in the collision operation. In the steady state it will therefore oscillate between two values and we have the self-consistency condition \(\left(\left<v_x v_y\right>-{\dot \gamma}{\delta t}\left<v_y^2\right>\right)f=\left<v_xv_y\right>\). Thus
\begin{eqnarray}
\left< v_x v_y \right> = - \frac{{\dot \gamma}{\delta t}f}{1-f}\left<v_y^2\right>\label{BALANCEEQN}. 
\end{eqnarray}
Substitution into (\ref{sigma}) gives
\begin{eqnarray}
\sigma_{xy} = \rho{\dot \gamma}{\delta t}\left< v_y^2 \right>\left( \frac{1}{2}  + \frac{f}{1-f} \right). 
\end{eqnarray}  
Using equipartition of energy, eqn. (\ref{f}) and the definition of viscosity (\ref{NONEQSV}) yields
\begin{eqnarray}
\eta^{2D}_{kin} = \frac{\gamma k_BT{\delta t}}{a^{2}} \left[ \frac{\gamma}{(\gamma-1+e^{-\gamma})(1-\cos2\alpha)} - \frac{1}{2}\right]\label{2dkinvistheory}.
\end{eqnarray}  
Applying the same proceedure in three dimensions gives
\begin{eqnarray}
&&\eta^{3D}_{kin}=\nonumber\\
&&\frac{\gamma k_BT{\delta t}}{a^3} \left[ \frac{5\gamma}{(\gamma-1+e^{-\gamma})(4 - 2 \cos\alpha - 2 \cos2\alpha) } - \frac{1}{2}\right]\label{finaletakin}.
\end{eqnarray}

%--------------------------------------------------------------------------

\subsection{Calculation of the collisional viscosity}
\par The collisional viscosity \(\eta_{col}\) can also be calculated using a kinetic theory approach. Consider a cubic cell with side \(a\). As in the case of the kinetic viscosity a shear rate \({\dot \gamma}=\frac{{\partial}u_x(y)}{{\partial}y}\) is applied along the \(x\)-axis. The system  is divided into two subcells by a plane at \(y=y_0\) where \(0\,{\leq}\,y_0\,{\leq}\,a\). Let the subcell at \(y_0\,{<}\,y\,{\leq}\,a\) contain \(n_1\) particles with an average collective flow velocity \(\mbox{\boldmath $u_1$}\) and that at \(0\,{<}\,y\,{\leq}\,y_0\) contain \(n_2\) particles with an average velocity \({\mbox{\boldmath $u_2$}}\). We have \(u_{1x}-u_{2x}=\frac{n_1+n_2}{n_2}\left(u_{1x}-v_{cmx}\right)\) where \(v_{cmx}\) denotes the \(x\)-component of the centre of mass velocity of the cell. As the average distance between the subcell velocities is \({\delta}y=\frac{a}{2}\) the shear rate is
\begin{eqnarray}
{\dot \gamma}=\frac{{\partial}u_x}{{\partial}y}&=&\frac{u_{1x}-u_{2x}}{{\delta}y}\nonumber\\
&=&\frac{2n}{a\left(n-n_1\right)}\left(u_{1x}-v_{cmx}\right)\label{SHEARRATECOL}
\end{eqnarray}
where \(n=n_1+n_2\) is the number of particles in the cell.

\par Our aim is to calculate the momentum crossing the plane at \(y_0\). We consider the \(n_1\) particles at \(y_0\,{<}\,y\,{\leq}\,a\). The momentum transfer between the two subcells
\begin{eqnarray}
\sigma_{xy}=-\left.\left(\sum_{i}p_{ix}(t+{\delta}t)-\sum_{i}p_{ix}(t)\right)\right/\left(a^{d-1}{\delta}t\right)
\end{eqnarray}
where \(i\) runs over the \(n_1\)-particles, is calculated using the collision operation (\ref{COLLISIONSTEP}). Averaging over an isotropic distribution of the rotation axis gives
\begin{eqnarray}
\sigma_{xy}=\frac{m}{a^{d-1}{\delta}t}\left[\frac{2}{d}n_1\left(1-\cos{\alpha}\right)\left(u_{1x}-v_{cmx}\right)\right]\label{STRESSCOL}
\end{eqnarray}
where \(m\) is the mass of the solvent particle. Using eqns. (\ref{SHEARRATECOL}), (\ref{STRESSCOL}) and the definition of the viscosity (\ref{NONEQSV}) we obtain
\begin{eqnarray}
\eta_{col}=\frac{mn_1\left(n-n_1\right)}{a^{d-2}dn{\delta}t}\left(1-\cos{\alpha}\right).\nonumber
\end{eqnarray}

\par As \(n_1+n_2\) is in general small we must again consider fluctuations in the particle density. The numbers in the subcells, \(n_1\) and \(n_2\) are binomially distributed and averaging over them gives
\begin{eqnarray}
\eta_{col}=\frac{m\left(1-\cos{\alpha}\right)}{a^{d-2}d{\delta}t}\left(n-1\right)\!\left(\frac{y_0}{a}\right)\!\!\left(1-\frac{y_0}{a}\right)\nonumber.
\end{eqnarray}
Here we used \({\sum}n_1\!\left(n-n_1\right)\!P(n_1)_{bin}=n^2p-npq-n^2p^2\) where \(p=1-\frac{y_0}{a}\) and \(q=\frac{y_0}{a}\). Using the Poisson distribution to average over $n$
\begin{eqnarray}
\eta_{col}&=&\frac{m\left(1-\cos{\alpha}\right)}{a^{d-2}d{\delta}t}\left(\frac{y_0}{a}\right)\!\!\left(1-\frac{y_0}{a}\right)\!\sum_{n=2}^{\infty}\left(n-1\right)\frac{e^{-\gamma}{\gamma}^n}{n!}\nonumber\\
&=&\frac{m\left(1-\cos{\alpha}\right)}{a^{d-2}d{\delta}t}\left({\gamma}-1+e^{-{\gamma}}\right)\!\left(\frac{y_0}{a}\right)\!\!\left(1-\frac{y_0}{a}\right)\nonumber
\end{eqnarray}
where the summation runs from $2$ to $\infty$ because there is no momentum transfer unless  \(n_1{\geq}1\) and \(n_2{\geq}1\). Finally, averaging over all planes \(0\,{\leq}\,y_0\,{\leq}\,a\) gives
\begin{eqnarray}
\eta_{col}=\frac{m\left(1-\cos{\alpha}\right)}{6a^{d-2}d{\delta}t}\left({\gamma}-1+e^{-{\gamma}}\right)\label{3DCOLVIS}.
\end{eqnarray}

\par In the large density limit the effect of the density fluctuations vanish and the theory agrees with the expression obtained by Ihle and Kroll\cite{IK01} for the two dimensional case (\(d=2\)). Also for case \(\alpha = \frac{\pi}{2}\) and \(d=2\) the result presented here is identical to that given in\cite{MK99} by Malevanets and Kapral. When \(d=3\) we find that our expression differs from that given in \cite{MK99} for \(\alpha=\frac{\pi}{2}\).
 
%--------------------------------------------------------------------------

\subsection{Measuring the shear viscosity numerically}

The computationally most efficient approach to measure the shear
viscosity directly exploits the experimental definition of
viscosity (\ref{NONEQSV}). A shear is applied along $x$ setting 
up a velocity gradient along $y$. The stress tensor \(\sigma_{xy}\), equal to the flux of \(x\)-momentum
across a plane perpendicular to the $y$ axis can easily be measured numerically.

\begin{figure}[hbp]
\begin{center}
\centerline{\epsfxsize=12cm \epsffile{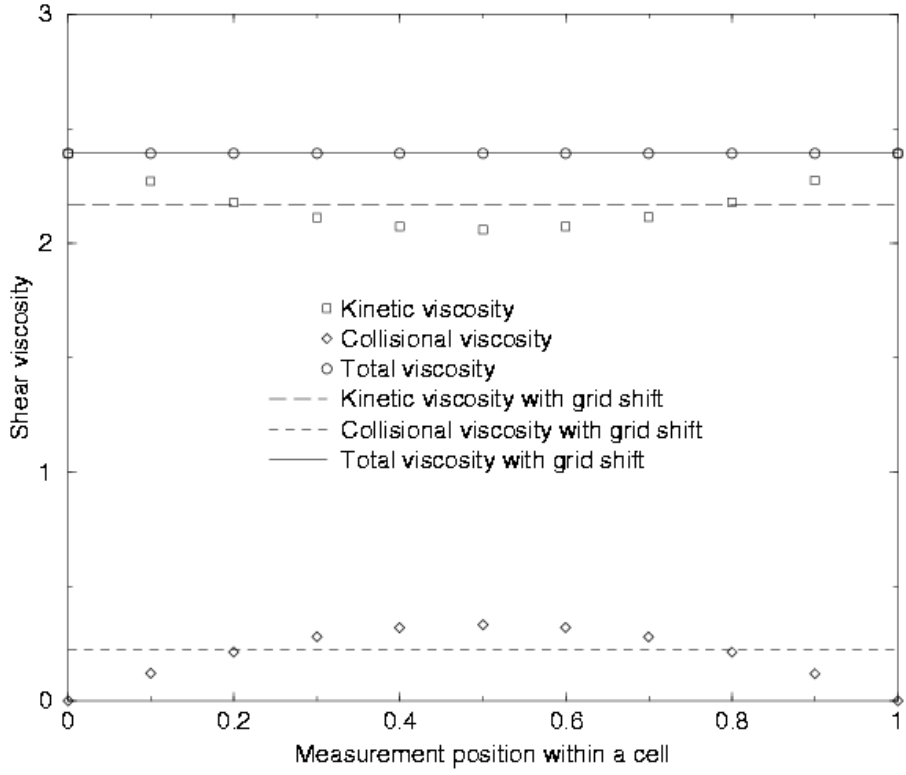} }
\end{center}
\caption{Dependence of the shear viscosity $\eta$ on the position of the measurement plane relative to a cell for \(k_BT=0.8\), \({\gamma}=5.0\) and \(\alpha=\frac{\pi}{2}\).}
\label{a01}
\end{figure}

Again we consider the momentum transfer in two parts, that due to the streaming step and that due
to the collision step. A subtlety that needs to be taken into account occurs because of the
finite size of the grid imposed in the collision step. This means that
the streaming and collisional contributions to the momentum transfer
depend on the position of the plane chosen for the measurement. For
example, if the plane lies at the edge of a cell, there is no momentum
transfer during the collision step whereas if the plane is taken at the
centre of a cell the momentum transfer in the collision is
maximised. The streaming contribution also varies, in such a way that
the total momentum transfer in one simulation time step is independent
of the position of the measuring plane.  This is illustrated in figure 
\ref{a01}. To separate out contributions to the kinetic and collisional     %---------Figure ref here for fig 1
viscosities it is necessary to average over all plane positions
within a cell: this is another advantage of imposing the grid shift
which performs the averaging over the cell automatically.

To measure the shear viscosity we used a simulation of size 
\(32\times32\times32\). Four thousand time steps were required to 
reach the steady state and the stress tensor was averaged over ten 
thousand time steps. 
It is also necessary to average over many measurement planes to
minimise thermal noise. We choose \(25\) planes from the \(32\) 
cubed box. Planes near the edge of the
simulation box are omitted to avoid edge effects resulting from the
Lees-Edwards boundary conditions. 
We also note that the pressure, \(P=-\sigma_{yy}\), can be measured in
the same way, and takes the expected value for an ideal gas.

\begin{figure}[hbp]
\begin{center}
\centerline{\epsfxsize=12cm \epsffile{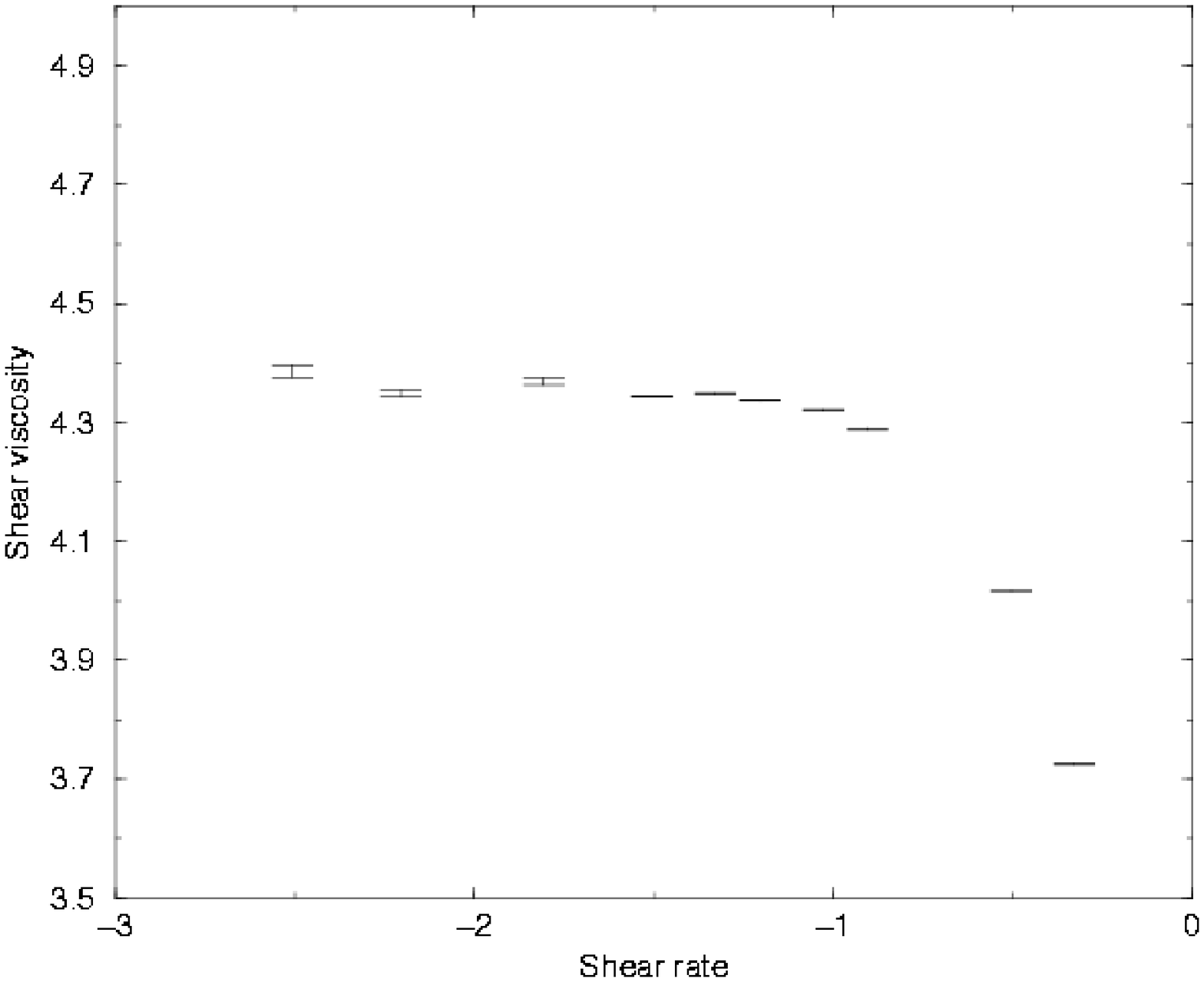} }
\end{center}
\caption{Dependence of the shear viscosity $\eta$ on shear rate \({\dot \gamma}\) for \(k_BT=0.8\), \(\gamma=5.0\) and \(\alpha=\frac{\pi}{3}\).}
\label{a02}
\end{figure}

To check that the algorithm corresponds to a Newtonian fluid we plot
in figure \ref{a02} the dependence of the shear viscosity on the shear rate.        %------Figure ref here for fig 2
The viscosity is independent of the shear rate, as expected, except
at very low and very high shear rates. The deviations at low shear are
due to statistical errors because
the shear represents only a small perturbation to the Maxwellian
velocity distribution. Those at high shear result from finite-size
effects. They occur when the distance moved by the walls in each time
step approaches the size of the system. The value of  
\({\dot {\gamma}}\) chosen to measure the shear viscosity was \(0.0625\) at \(k_BT=0.8\).

The shear viscosity  \({\eta}_{GK}\) can also be measured using an equilibrium 
approach based on the Green-Kubo formula\cite{MK00,IK01,K78}
\begin{eqnarray}
{\eta}_{GK}=\lim_{t{\rightarrow}\infty}\frac{V}{k_BT}\int_0^tdS
\left<{\sigma}_{xy}(S){\sigma}_{xy}(0)\right>\!. \label{DEGKFSV}
\end{eqnarray}
For discrete time steps
this can be approximated by  
\begin{eqnarray}
&&{\eta}_{GK}=\nonumber\\
&&\frac{V{\delta}t}{k_BT}\left(\frac12\left<{\sigma}_{xy}(0){\sigma}_{xy}(0)\right>+\sum_{n=1}^{\infty}\left<{\sigma}_{xy}(n{\delta}t){\sigma}_{xy}(0)\right>\right)
\end{eqnarray}
where \({\delta}t\) is the time step of the solvent dynamics.

Further details of the equilibrium approach is described in \cite{IK01}. However thermal fluctuations mean that very long
runs are needed to obtain statistically meaningful results -- using
shear is faster by a factor $\sim {30}$. Therefore
this approach was used only as a check on the validity of the
non-equilibrium method.

\subsection{Results for the viscosity}

\par
\begin{figure}[hbp]
\begin{center}
\centerline{\epsfxsize=11cm \epsffile{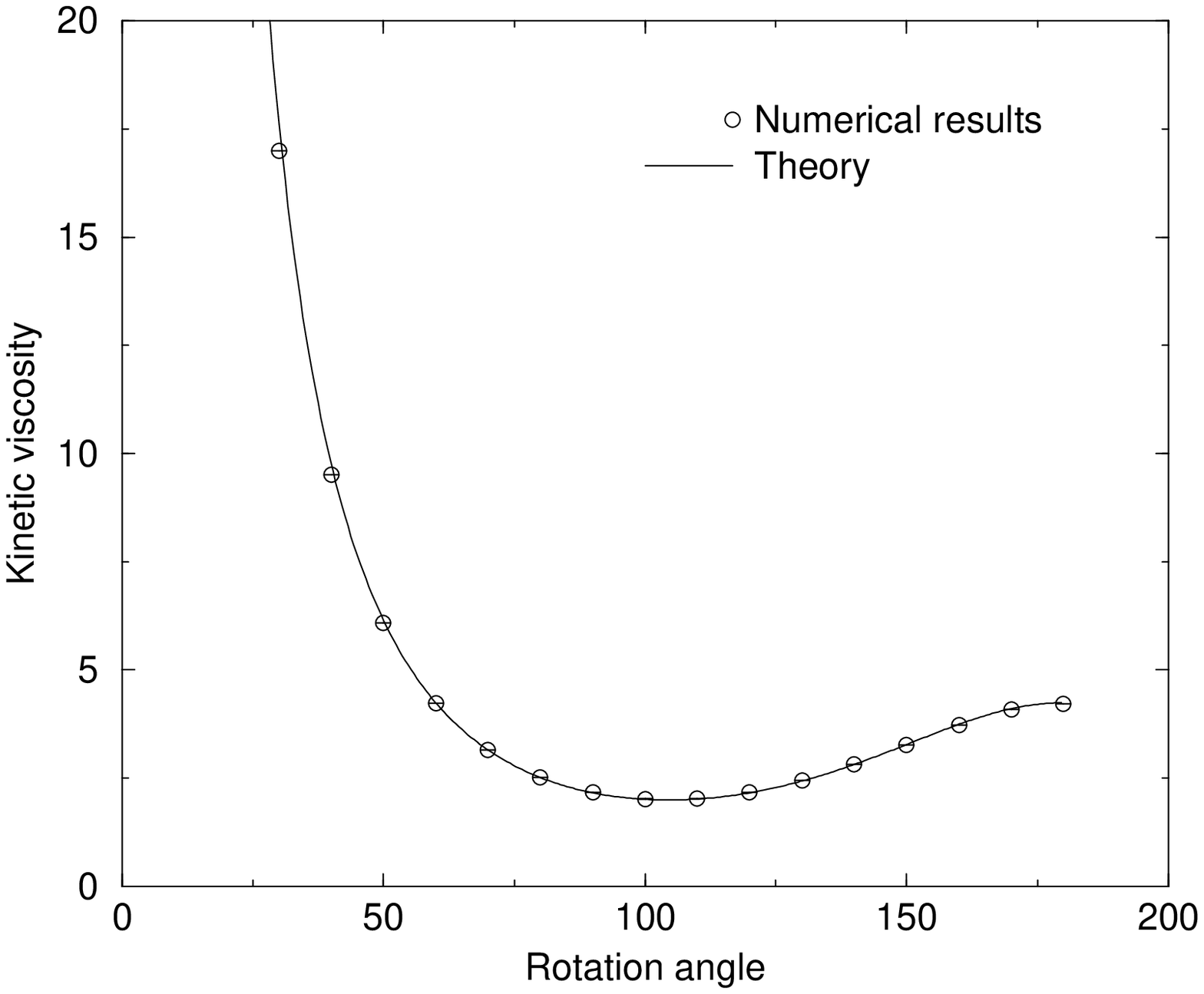} }
\end{center}
\caption{Dependence of the kinetic viscosity $\eta_{kin}$ on rotation angle $\alpha$ for \(k_BT=0.8\) and \(\gamma=5.0\).}
\label{a03}
\end{figure}

\begin{figure}[hbp]
\begin{center}
\centerline{\epsfxsize=11cm \epsffile{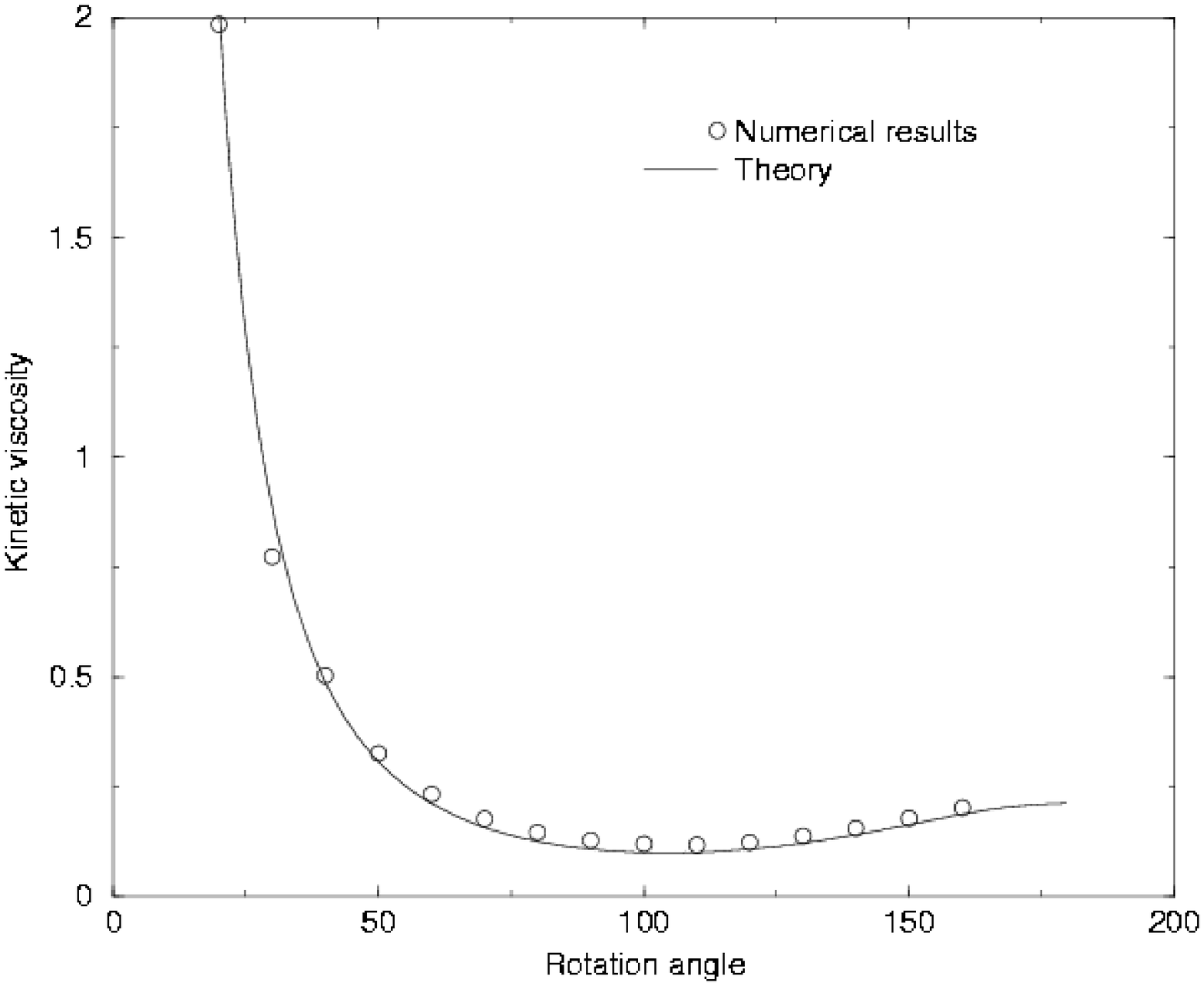} }
\end{center}
\caption{Dependence of the kinetic viscosity $\eta_{kin}$ on rotation angle $\alpha$ for \(k_BT=0.04\) and \(\gamma=5.0\).}
\label{a04}
\end{figure}

 Consider first the kinetic viscosity. Figures \ref{a03} and \ref{a04} present the dependence      %----------Figure ref here for fig 3 and 4 
of the three dimensional kinetic viscosity on the rotation angle \(\alpha\) at 
two different temperatures. Agreement between the theoretical result eqn (\ref{finaletakin})  
and the simulations is excellent.
As \(\alpha\) goes to zero the kinetic viscosity diverges. This corresponds to particles
moving with an infinite mean free path, unimpeded by collisions. From eqn
(\ref{f}) and (\ref{BALANCEEQN}), as \(\alpha\rightarrow0\) the
correlation \(\left<v_xv_y\right>\rightarrow-\infty\). In two dimensions the kinetic viscosity is symmetrical about \(\alpha = \frac{\pi}{2}\) and hence also diverges at \(\alpha = \pi\). This differs from the three dimensional case where the viscosity reaches a local maximum at \(\alpha = \pi\).

\par
\begin{figure}[hbp]
\begin{center}
\centerline{\epsfxsize=11cm \epsffile{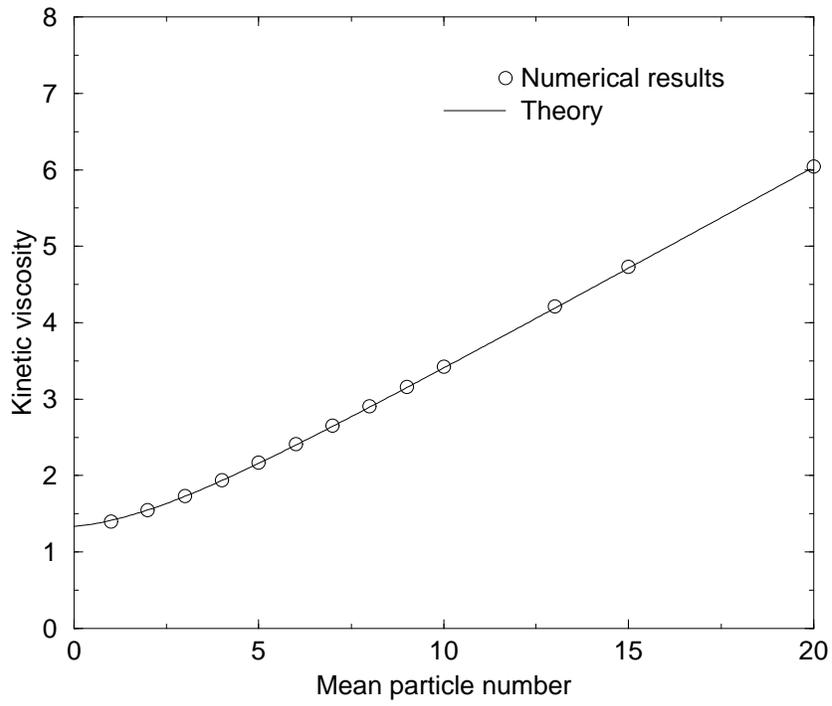} }
\end{center}
\caption{Dependence of the kinetic viscosity $\eta_{kin}$ on mean particle number $\gamma$ for \(k_BT=0.8\) and \(\alpha=\frac{\pi}{2}\).}
\label{a05}
\end{figure}

 A small deviation between the theory and the numerical results is observed 
at low temperatures. We believe that this is due to one of two effects. At such 
low temperatures, even with a gridshift, the assumption of molecular chaos is an approximation. Also at these temperatures the equilibrium distribution no longer 
approximates to a Maxwell-Boltzmann distribution due to the effect of the shear. Figure \ref{a05} compares the density dependence of the kinetic viscosity with theory, again showing excellent agreement. 

\par
\begin{figure}[hbp]
\begin{center}
\centerline{\epsfxsize=11cm \epsffile{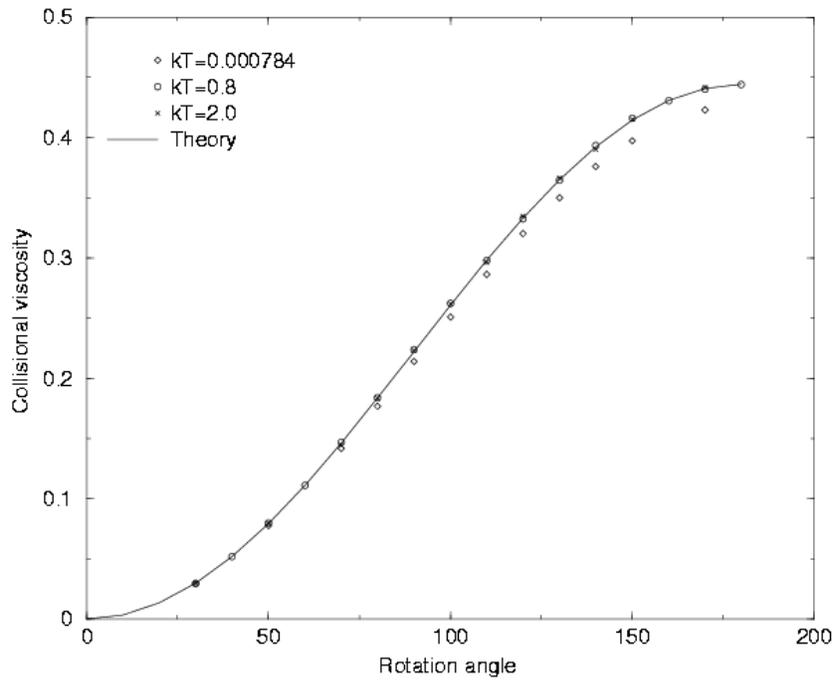} }
\end{center}
\caption{Dependence of the collisional viscosity $\eta_{col}$ on rotation angle $\alpha$ for \(\gamma=5.0\).}
\label{a06}
\end{figure}

\begin{figure}[hbp]
\begin{center}
\centerline{\epsfxsize=11cm \epsffile{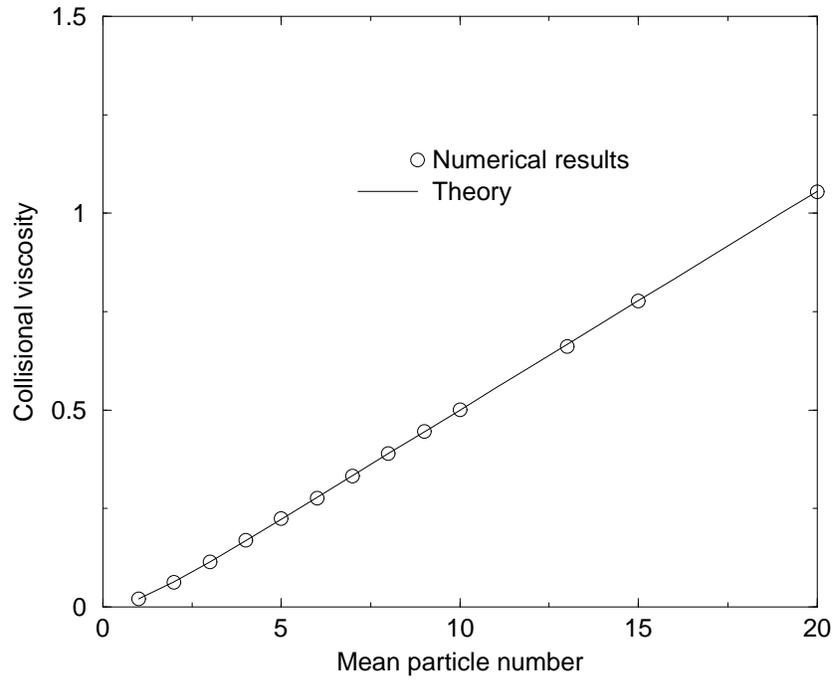} }
\end{center}
\caption{Dependence of the collisional viscosity $\eta_{col}$ on mean particle number $\gamma$ for \(k_BT=0.8\) and \(\alpha=\frac{\pi}{2}\).}
\label{a07}
\end{figure}

 We now turn to the collisional viscosity. This is as predicted, independent 
of temperature (eqn (\ref{3DCOLVIS})). Figures \ref{a06} and \ref{a07} show its dependence on rotation angle \(\alpha\) and particle density respectively. Agreement with the theory is again excellent 
except at extremely low temperatures.

\par The kinetic viscosity is proportional to the temperature and we
recall that the collisional viscosity does not depend on the temperature.
Hence the former dominates at high temperatures, and the latter at low
temperatures. This has important indications for the balance of viscous and diffusive transport as discussed below.

\par
\begin{figure}[hbp]
\begin{center}
\centerline{\epsfxsize=12cm \epsffile{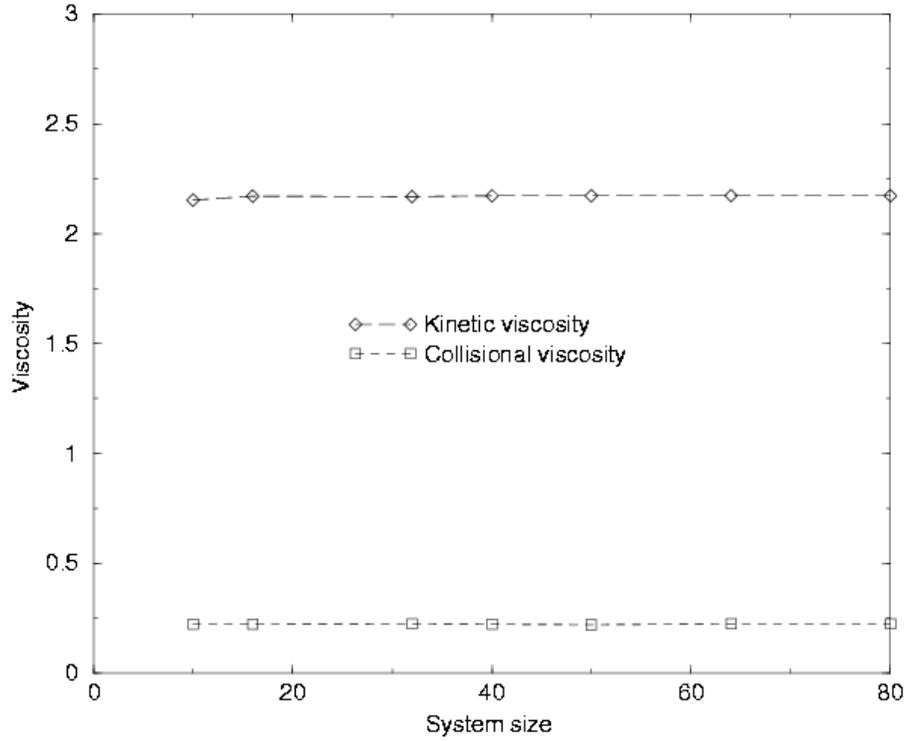} }
\end{center}
\caption{Effect of finite lattice size on the measured viscosity for $k_BT=0.8$, $\gamma=5$ and \(\alpha=\frac{\pi}{2}\).}
\label{a08}
\end{figure}

 Finally we check for any finite size effects in the simulations. Figure \ref{a08} shows that for the systems studied there is only a very small effect on the viscosity at the smallest 
system size.

%--------------------------------------------------------------------------

\section{Coupling to a solute particle}
\par To place solute particles such as polymers or colloids in the stochastic rotation dynamics solvent it is important to understand the coupling between them and the solvent. Therefore we calculate here the friction coefficient acting on a particle of mass \(M\) and velocity \({\mbox{\boldmath $v$}}\) in a Brownian solvent, and then show that the fluctuation-dissipation theorem holds as expected.

\subsection{Friction coefficient}
\par The discretized Langevin equation for a monomer is
\begin{eqnarray}
M\frac{{\mbox{\boldmath $v$}}(t\!+\!{\delta}t)-{\mbox{\boldmath $v$}}(t)}{{\delta}t}=-\xi{\mbox{\boldmath $v$}}(t)+{\mbox{\boldmath $\eta$}}(t)\label{DISLANGEVIN}
\end{eqnarray}
where \(\xi\) denotes the friction and  $\delta t$ is the time step. \({\mbox{\boldmath $\eta$}}(t)\) is stochastic noise with time average \(\left<{\mbox{\boldmath $\eta$}}\right>=0\). The evolution of the velocity is given by eqn (\ref{COLLISIONSTEP}) which may be rewritten
\begin{eqnarray}
{\mbox{\boldmath $v$}}(t\!+\!{\delta}t)-{\mbox{\boldmath $v$}}(t) = \left({\mbox{\boldmath $R$}}-{\mbox{\boldmath $I$}}\right)\left({\mbox{\boldmath $v$}}(t)-{\mbox{\boldmath $v_{cm}$}}\right)\label{MKLANGEVIN}
\end{eqnarray}
where \({\mbox{\boldmath $R$}}\) is the usual rotation matrix and \({\mbox{\boldmath $I$}}\) is the identity matrix.

\par We consider a particle of mass $M$ in a cell with \(n\) solvent particles of mass \(m\). The centre of mass velocity of the cell is given by
\begin{eqnarray}
{\mbox{\boldmath $v_{cm}$}}=\frac{M{\mbox{\boldmath $v$}}(t)+\sum_{i=1}^nm{\mbox{\boldmath $u_i$}}}{M+mn}.\label{VCM}
\end{eqnarray}
Averaging eqn (\ref{MKLANGEVIN}) over an isotropic distribution of the rotation axis and over $\mbox{\boldmath $u_i$}$, assuming $\left< \mbox{\boldmath $u_i$} \right> = 0$, gives
\begin{eqnarray}
\left< {\mbox{\boldmath $v$}}(t\!+\!{\delta}t)-{\mbox{\boldmath $v$}}(t) \right>
= \frac{2}{3}\frac{mn}{M+mn}\left(1-\cos{\alpha}\right) \left< {\mbox{\boldmath $v$}}(t) \right>.
\label{vv}
\end{eqnarray}
In time the number of particles within a given cell, $n$, changes. The probability is given by the Poisson distribution, $P(n) = \frac{e^{-\gamma}{\gamma}^n}{n!}$, where $\gamma$ is the average number per cell. By comparing (\ref{vv}) with the time averaged Langevin equation (\ref{DISLANGEVIN}), we identify the friction coefficient as
 
\begin{eqnarray}
\xi^{3D} = \sum_{n=0}^{\infty} \frac{e^{-\gamma}{\gamma}^n}{n!}\frac{mn}{M+mn} \frac{2M}{3\,{\delta}t} \left(1-\cos{\alpha}\right).\label{3DFRIC}
\end{eqnarray}
In two dimensions
\begin{eqnarray}
\xi^{2D} = \sum_{n=0}^{\infty} \frac{e^{-\gamma}{\gamma}^n}{n!}\frac{mn}{M+mn} \frac{M}{\,{\delta}t} \left(1-\cos{\alpha}\right).\label{2DFRIC}
\end{eqnarray}

\subsection{Numerical results for the friction coefficient}

\par
\begin{figure}[hbp]
\begin{center}
\centerline{\epsfxsize=11cm \epsffile{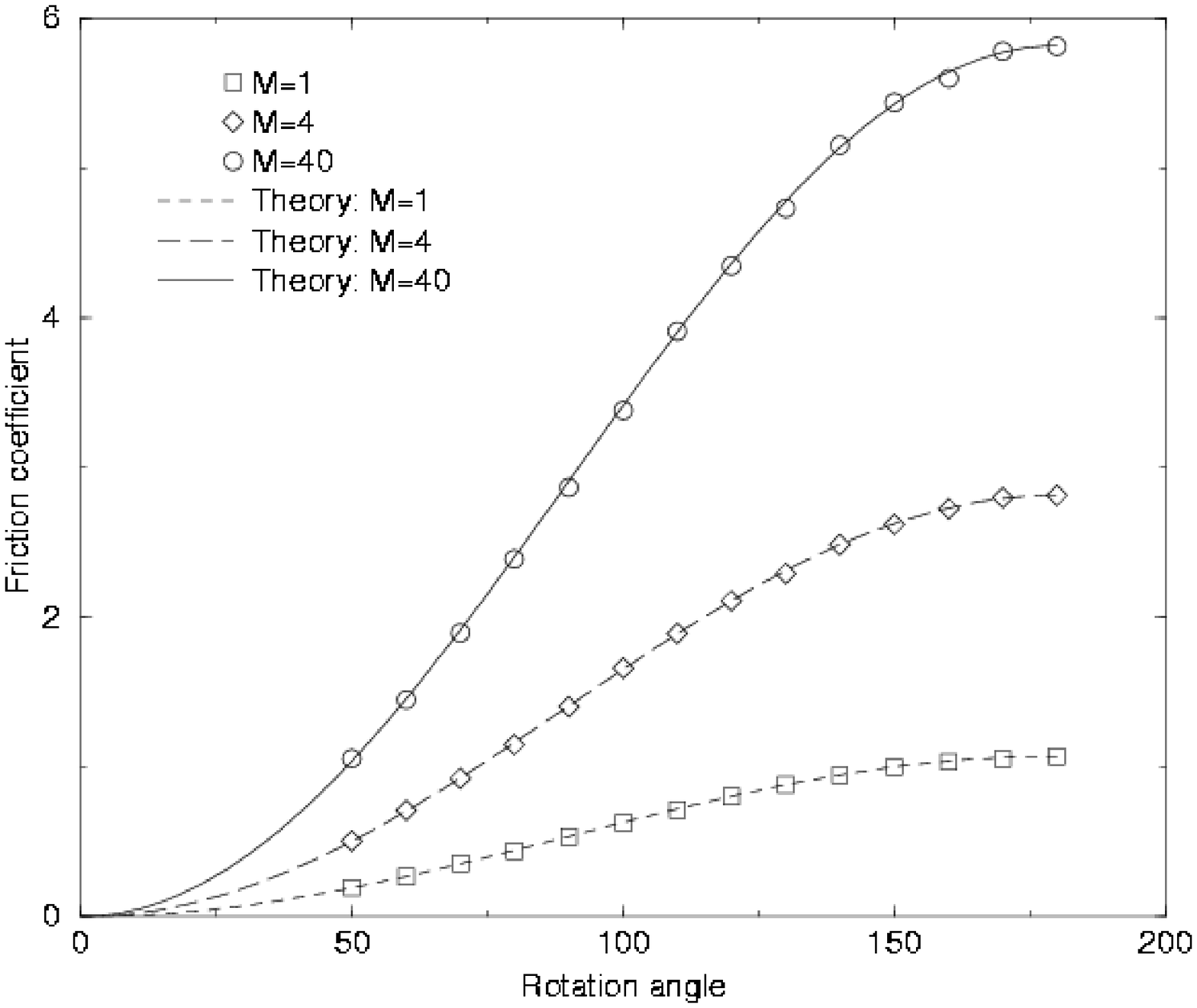} }
\end{center}
\caption{Dependence of the friction coefficient $\xi$ on rotation angle $\alpha$ for different particle masses \(M\) for \(k_BT=0.8\) and \(\gamma=5.0\).}
\label{a09}
\end{figure}

 The friction coefficient is measured numerically using eqn (\ref{DISLANGEVIN}). Multiplying  (\ref{DISLANGEVIN}) by \({\mbox{\boldmath $v$}}(t)\) and taking a time average 
\begin{eqnarray}
\left<{\mbox{\boldmath $v$}}(t\!+\!{\delta}t)\!\cdot\!{\mbox{\boldmath $v$}}(t)\right>
-\left<{\mbox{\boldmath $v$}}^2(t)\right>=-\frac{{\delta}t}{M}\xi\left<{\mbox{\boldmath $v$}}^2(t)\right>
\label{vcor}
\end{eqnarray}
where we use \(\left<{\mbox{\boldmath $\eta$}}(t)\!\cdot\!{\mbox{\boldmath $v$}}(t)\right>=\left<{\mbox{\boldmath $\eta$}}(t)\right>\!\left<{\mbox{\boldmath $v$}}(t)\right>=0
\). Therefore, using the equipartition theorem
\begin{eqnarray}
\xi=\frac{M}{{\delta}t}\left(1-\frac{M}{dk_BT}\left<{\mbox{\boldmath $v$}}(t\!+\!{\delta}t)\!\cdot\!{\mbox{\boldmath $v$}}(t)\right>\right)\label{NUMFRIC}
\end{eqnarray}
Theory and numerical results are in good agreement for different \(M\) as shown in figure \ref{a09}.

\subsection{Fluctuation-dissipation theorem}
\par Next we verify fluctuation-dissipation theorem. Substituting (\ref{MKLANGEVIN}) into (\ref{DISLANGEVIN}) and rearranging gives
\begin{eqnarray}
{\mbox{\boldmath $\eta$}}(t) = \!\!\frac{M}{{\delta}t}\!\!\left({\mbox{\boldmath $R$}}-{\mbox{\boldmath $I$}}\right)\left({\mbox{\boldmath $v$}(t) - \mbox{\boldmath $v_{cm}$}}\right) + \xi {\mbox{\boldmath $v$}}(t).  \label{NOISE01}
\end{eqnarray}
To calculate the equal time correlation function we must assume equipartition of energy and molecular chaos, therefore $\left< {\mbox{$v_{\alpha}$}}{\mbox{$u_{i \beta}$}} \right> = \left< {\mbox{$v_{\alpha}$}} \right> \left< {\mbox{$u_{i \beta}$}} \right> = 0$ and  $\left< {\mbox{$u_{i \alpha}$}}{\mbox{$u_{j \beta}$}} \right> = \delta_{i j} \delta_{\alpha \beta} k_B T/m$. Using eqn (\ref{VCM}) and averaging over an isotropic distribution of the rotation axis and the solvent velocity distribution we find
\begin{eqnarray}
\left<{\mbox{\boldmath $\eta$}}(t)\!\cdot\!{\mbox{\boldmath $\eta$}}(t)\right>&=&\frac{2d\,{\xi}k_BT}{{\delta}t}\left(1-\frac{mn\left(1-\cos{\alpha}\right)}{d\left(M+mn\right)}\right)
\end{eqnarray}
for dimension $d$. The first term in the bracket is that obtained from the fluctuation-dissipation theorem. The second is a small correction, which goes to zero in the limit of large $M$.

\subsection{Diffusion coefficient}
The diffusion coefficient can be calculated using the definition

\begin{eqnarray}
D&=&\lim_{t{\rightarrow}\infty}\frac{\left<{\left(\mbox{\boldmath $r$}\left(t\right)-\mbox{\boldmath $r$}\left(0\right)\right)}^2\right>}{2dt}\nonumber\\
&=&\frac{1}{2d}\left<\mbox{\boldmath $v$}\left(0\right)\cdot\mbox{\boldmath $v$}\left(0\right)\right>{\delta}t
+\frac{1}{d}\sum_{n=1}^{\infty}\left<\mbox{\boldmath $v$}\left(n \delta t\right)\cdot\mbox{\boldmath $v$}\left(0\right)\right>{\delta}t.\label{DGK01}
\end{eqnarray}
We note that the resulting equation (\ref{DGK01}) is the same as the Green-Kubo formula, which is derived from linear response theory\cite{K78}.

\par For a Brownian particle\cite{note1} the velocity-velocity autocorrelation is easily calculated using eqn (\ref{vcor})
\begin{eqnarray}
\left<{\mbox{\boldmath $v$}}(t\!+\!n{\delta}t)\!\cdot\!{\mbox{\boldmath $v$}}(t)\right>={\left(1-\frac{{\delta}t}{M}\xi\right)}^{\!n}\!\left<{\mbox{\boldmath $v$}}^2(t)\right>.\label{DGK02}
\end{eqnarray}
Substitution into (\ref{DGK01}) gives,
\begin{eqnarray}
D=\frac{k_BT}{\xi}\left(1- \frac{\delta t \xi}{2M} \right).\label{DGK04}
\end{eqnarray}
This is the normal Einstein relation with an error term which becomes vanishingly small in the large $M$ limit.

%--------------------------------------------------------------------------

\section{Conclusion}

We have provided analytic expressions for the shear viscosity of a two- and three-dimensional mesoscale model of hydrodynamics, stochastic rotation dynamics. The derivation assumes molecular chaos. The viscosity has two contribution $\eta_{kin}$ and $\eta_{col}$ from the streaming and collision steps of the algorithm respectively. The former dominates at high temperatures and the latter at low temperatures.

We also obtained an expression for the viscous drag \(\xi\) acting on a test particle. Note that $\xi$ and $\eta_{col}$ given by eqns (\ref{3DFRIC}) and (\ref{3DCOLVIS}) have the same dependence on the rotation angle $\alpha$ and are both independent of temperature. This is to be expected as they both result from momentum transfer in the collision step of the algorithm. It is useful to realise that in the stochastic rotation method for a fluid with a given total viscosity the coupling to a test particle will depend on temperature and will be very weak at high temperatures where \(\eta_{kin}\gg\eta_{col}\). 

The viscosity in three dimensions was measured numerically by using Lees-Edwards boundary conditions to apply a shear to the system. The analytic and numerical results were found to be in excellent agreement thus providing a check on both calculations. It is important to have a robust way of measuring transport coefficients for this model as it can easily be coupled to a solute such as polymers or colloids where the non-Newtonian flow properties of the resultant solution are of interest.
%--------------------------------------------------------------------------

%--------------------------------------------------------------------------

\end{multicols}

%--------------------------------------------------------------------------

%--------------------------------------------------------------------------

\begin{references}
\bibitem{note1} The self-diffusion coefficient of a single monomer is measured in ref \cite{KGY02} as ($D_H=0.146$ (hydrodynamic) and $D_B=0.137$ (Brownian)). The small change is due to the hydrodynamic long time tail.
\end{references}
\end{document}